\documentclass[aip,rsi,amsmath,amssymb,article]{revtex4-1}
\usepackage{graphicx}
\usepackage[utf8x]{inputenc}

\usepackage{color}           % Allows use of font color
\usepackage{colortbl}  %Allows use of color in tables

\begin{document}

\title{Ultraviolet stimulated electron source for use with low energy plasma instrument calibration}

\author{Kevin Henderson\footnote{Corresponding author: kch@lanl.gov}, Ron Harper, Herb Funsten, and Elizabeth MacDonald}
%\email[Corresponding author: ]{kch@lanl.gov}
\affiliation{Space Science and Applications, Los Alamos National Laboratory, Los Alamos, New Mexico}

\pacs{07.77.Ka, 84.60.Jt, 61.80.Ba, 68.43.Tj, 65.40.gh}
\date{\today}

\begin{abstract}
We have developed and demonstrated a versatile, compact electron source that can produce a monoenergetic electron beam up to 50 mm in diameter from 0.1 to 30 keV with an energy spread of $<$10 eV. By illuminating a metal cathode plate with a single near ultraviolet (UV) light emitting diode (LED), a spatially uniform electron beam with 15\% variation over 1 cm$^2$ can be generated. A uniform electric field in front of the cathode surface accelerates the electrons into a beam with an angular divergence of $<$1$^\circ$ at 1 keV. The beam intensity can be controlled from $10-10^9$ electrons cm$^{-2}$ s$^{-1}$.\\

%{\color{darkblue} Potential referees\\
%Frederico Allegrini, Walter Gekelman, Lessard Marc (UV NH), Markus Shapirio (GSFC), Sarah Jones }
\end{abstract}

\maketitle

% #########################################################################
\section{Introduction}

The production and control of spatially broad electron beams with energies up to tens of kV and fluxes up to $10^9$ electrons/cm$^2$/s has many important applications, such as characterizing the performance of radiation detectors and spectrometers, electron spectroscopy, charge management, and electron diffraction.\cite{SSD1,SSD2, Plasma0,Plasma1,Plasma2, Katrin1, LISA1,LISA2} Standard types of electron sources are typically used for these purposes, although they often have limitations that increase their complexity for use or constrain their utility. For example, beta sources generate electrons from radioactive decay, are readily available, and provide stable, predictable electron fluxes; however, they generate a broad spectrum of electron energies, are potentially hazardous, are often accompanied by gamma ray emission, and are limited in flux and energy range. Traditional thermionic emission cathodes can generate high electron fluxes ($>$ 10 nA) but typically have poor dynamic control, tend to be expensive, and are usually employed for generation of narrow beams with limited energy ranges.\cite{Kimball} State-of-the-art carbon nanotube cathodes are robust field emitters that tend not to degrade with time,\cite{CNT} but operation of the source can be complex. 

In contrast, electron sources based on photoelectrons generated by stimulation of a photocathode with ultraviolet (UV) light can be comparatively straightforward, both in design and operation.\cite{Plasma0} Ultraviolet light sources such as xenon, mercury, and deuterium lamps are widely available with relatively high power. Importantly, they emit UV over a broad, continuous spectral range, from near UV to far UV ($122 -400$ nm), which spans the work functions of most cathode materials. Therefore, we expect abundant photoelectron emission from these photocathodes mostly independent of the type of UV source. Traditional disadvantages of photoelectron-stimulated electron sources using these lamps is their poor dynamic control, large volume requirements, and intrinsic inefficiency that drives high power requirements and simultaneous production of copious but unusable background light. Recently, however, significant progress of solid state technologies that produce UV light have demonstrated higher efficiency (output UV photon per unit input power), increased output UV flux, and have pushed to shorter UV wavelengths. UV laser systems, typically diode pumped solid state (DPSS) lasers are still expensive and bulky and cannot generate continuous wave UV at wavelengths lower than 375 nm. However, the last five years has seen a dramatic advance in the availability and performance of inexpensive UV light emitting diodes (LEDs) at wavelengths $<375$ nm. Not surprisingly, UV LED stimulated electron sources have recently been demonstrated for use with precision electron spectroscopy for the KATRIN neutrino experiment\cite{Katrin1,Katrin2} as well as low resource, precision control charge management systems for the freely floating proof masses that could be used, for example, in the Laser Interferometer Space Antenna (LISA) mission.\cite{LISA1,LISA2}

Our interest in a low cost, highly controllable, mono-energetic, broad-beam electron source is driven by the need for characterization and calibration of space plasma and electron spectrometers as well as characterizing the performance of new low energy electron detectors and detection technologies. Furthermore, the source cannot produce copious gamma rays or high fluxes of background light that can generate background counts in detectors typically used in these spectrometers, such as microchannel plate (MCP) and channel electron multiplier (CEM) detectors. We have developed and demonstrated an electron source that generates photoelectrons using a UV LED that meets our design goals and provides an extremely cost-effective alternative to commercially available electron sources with similar capabilities. We expect that this type of electron source has broad utility for scientific research, characterization of radiation detectors and detection technologies, and materials analysis such as electron spectroscopy.

% #########################################################################
\section{Experimental setup}

Our objectives for this electron source were to generate a large-area, spatially uniform, monoenergetic electron beam with an energy range of 0.1 to 30 keV while minimizing the electron source volume and power requirements. The geometry of the electron source was driven by high voltage stand-off (up to 30 kV between electrodes) and electro-optic considerations such as effects of fringe fields on angular divergence of the electron beam. The photocathode material was chosen to be machined easily, be operated without special or in-situ preparation or conditioning, and produce a long-term stable electron beam.

The vacuum chamber used for calibration of flight plasma and electron spectrometers is typically operated at $10^{-8}$ torr, at which pressure the residual gases in the chamber can rapidly form adsorbed layers over all surfaces. This precludes the use of special, low work function photocathodes that must operate at UHV or must be continuously conditioned, for example by heating. However, the presence of adsorbed gases affects materials with higher work functions, usually making the work functions lower\cite{Yamamoto}, therefore the list of possible photocathode materials increases.  Importantly, the key innovation that enables photoelectron production at higher work function is the availability of UV LEDs at sufficiently short wavelengths.  Fig.\;\ref{fig:Schematic}  shows the basic apparatus which is described from left to right in the following sections. A photograph of the completed electron source is shown in Fig.\;\ref{fig:Photo}.

\subsection{Electron source apparatus}

The electro-optic design for acceleration of the electron beam is inherently simple, with only two requisite components to generate a uniform electric field for electron acceleration: the photocathode biased to the accelerating voltage and a grounded grid. The biased photocathode and grounded grid plate were mounted to a grounded ring with appropriate stand-off insulators. The grid plate, at ground potential during normal operation, was attached to the grounded ring using eight alumina standoffs so that the effects of biasing the grid relative to the ground potential of the vacuum chamber could be tested.  The diameter of the grid plate and mounting ring is larger than the diameter of the photocathode so that the apparatus can be conductively enclosed around its diameter to prevent stray electric fields from encroaching into the chamber and to reduce scattered light.

\begin{figure}[htp]
\begin{center}
\includegraphics[width=4in,height=3.19in]{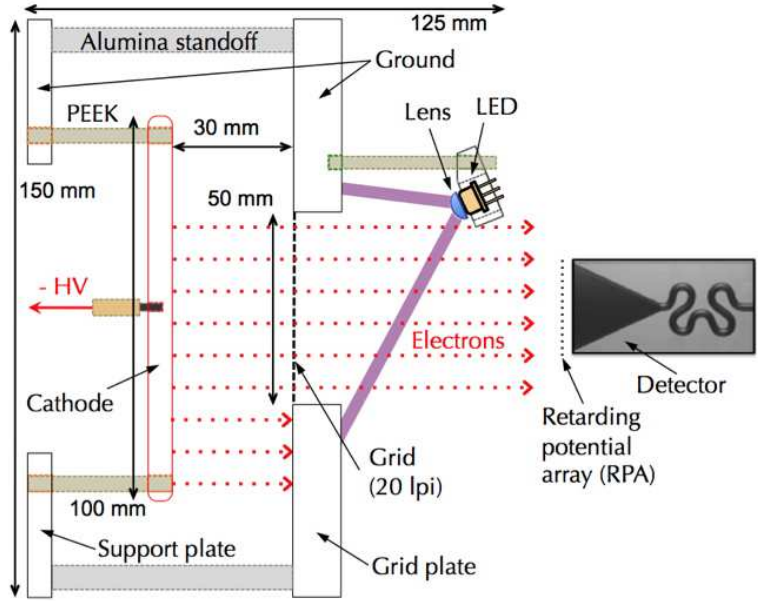}
\caption{(color online) Schematic showing the cross-section geometry of the apparatus.  The source lies within an envelope of $15 \times 15 \times 12.5$\;cm.  Except for the placement of the LED, the apparatus is cylindrically symmetric about the high voltage connector at the center of the cathode plate.  The LED lens is used to minimize divergence of the light, reducing scattered light that acts as a background in our detectors and enabling higher UV flux on the photocathode (and therefore increasing the output electron beam current). Note that multiple LEDs could be used to increase the output electron beam current, and that their mounting can be from the edge of the grid plate or on the face of the grid plate, as shown.  The highest beam flux we measured was $10^9$ electrons cm$^{-2}$ s$^{-1}$.}
\label{fig:Schematic}
\end{center}
\end{figure}

The cathode plate is $100$\;mm in diameter and $6.35$\;mm thick with radiused edges to reduce field emission. It is connected to the support plate with four insulating PEEK 6-32 threaded rods. The acceleration voltage (up to -30\;kV) is attached to the rear of the cathode plate. High voltage conditioning is performed prior to installation of the UV LED and uses the same negative high voltage power supply that is used to power the cathode plate during operation. Conditioning can be performed in air or in high vacuum since the shortest distances between cathode and any ground exceeds the minimum distance for dielectric breakdown of both air and vacuum by over a factor of two ($\sim 3$\;kV/mm for air and $>10$\;kV/mm in high vacuum). We found that high voltage conditioning of the photocathode is necessary to prevent high voltage microdischarges that could damage the UV LEDs, which are sensitive to electrostatic discharge (ESD). Furthermore, conditioning also prevents the production of electron bursts associated with microdischarge that can become part of the output electron beam. Placing a low pass filter ($200\;\Omega$ resistor, $0.01\;\mu$F capacitor) on the LED input within $1$\;cm of the LED prevents premature failures of LEDs due to electrical transients induced by microdischarges.

\begin{figure}[htp]
\begin{center}
\includegraphics[width=3.5in,height=2.97in]{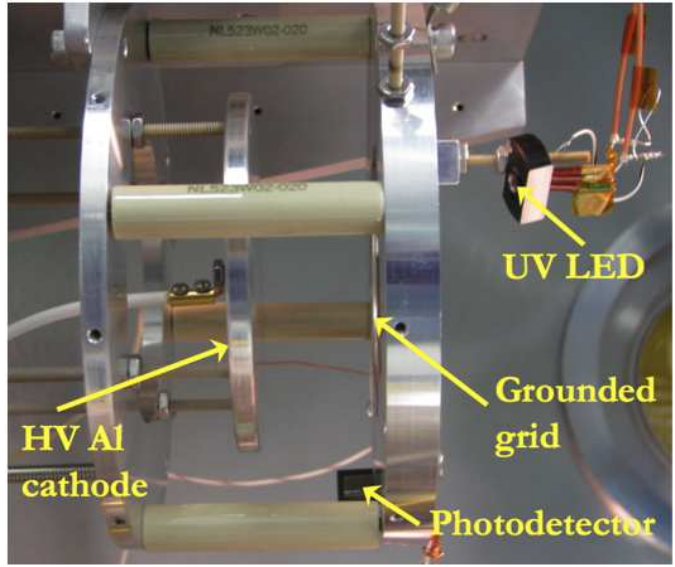}
\caption{(color online) Photograph of the UV LED stimulated electron source described in Fig.\;\ref{fig:Schematic}. The photodetector collects a portion of the diffuse reflected light off the cathode and is used to monitor for light output from the LED.}
\label{fig:Photo}
\end{center}
\end{figure}

The center of the grid plate has a $50$\;mm diameter hole. A hand-stretched, $20$\;line-per-inch (lpi) tungsten grid (92\% transmissive) is fastened to the grid plate with conductive tape. One or multiple LEDs can be mounted on the face of the grid plate, illuminating the cathode through the 50 mm diameter hole as shown in Fig.\;\ref{fig:Schematic}. The central axis of the LED is mounted at a small angle ($\sim 25^{\circ}$) relative to the electron beam direction, maximizing the UV flux on the region of the cathode that generates photoelectrons which are subsequently accelerated through the hole in the ground plate. The mounting geometry, number of LEDs, and LED lens properties must be carefully optimized and tested by optically imaging the projection of LED light onto the cathode imaging to ensure homogeneously illumination of the cathode.

\subsection{UV LED and Cathode}

Several types of low cost, low power, vacuum compatible UV LEDs are commercially available to stimulate photoelectrons from oxidized metal surfaces in a vacuum as poor as $10^{-6}$ torr. We tested UV LEDs at nominal output wavelengths of 260 nm (LED260W, Thorlabs), 295 nm (UVTOP295TO39FW, Sensor Electronic Technology), 310 nm (UVTOP310TO39FW, Sensor Electronic Technology), 365 nm (LED365W, Thorlabs), and 370 nm (LED370E, Thorlabs). Most results described here were obtained using Thorlabs LEDs at 260 nm and 365 nm. 

%We find that background counts in microchannel plate (MCP) and channel electron multiplier (CEM) detectors are substantially reduced or eliminated at higher wavelengths in the near UVA (320 - 400 nm) range, although the number of candidate photocathode materials with sufficiently high photoelectron production is restricted. 

For this study, all cathode materials were polished with 320 grit silicon carbide sandpaper and cleaned with isopropyl alcohol. Since no further surface preparation was employed, the surfaces rapidly oxidized and were susceptible to adsorbate accumulation in vacuum. The work functions of oxidized metals can range from 2.5 to 6 eV, corresponding to a wavelength range of 500 to 200 nm. Aluminum alloys, like 6061 (0.8-1.2\% Mg), which are inexpensive, readily available, and easily machined, have a 4.0 - 4.2 eV (290 -310 nm) work function for the bare metal surface.\cite{WFAl00}  The stable oxide layer that forms on the surface of the metal shifts the work function to a slightly lower energy of $\sim$ 3.6 - 3.9 eV (344 -318 nm). \cite{AlOxide0}

In practice, the work function of the surface depends on vacuum pressure (adsorbate composition and coverage), the presence of UV light, and on the preparation process. \cite{UVDesorp}  Interestingly, water adsorption can shift the bare metal work function lower, but the effect depends strongly on both the pressure of the system and UV light exposure \cite{AlOxideWater0,AlOxideWater1}. By monitoring the UV output flux over time and the electron current over time, we observe a reduction in the output efficiency (output electrons per incident UV photon) of the system for all photocathodes tested.  Using a 370 nm LED we observed photoelectron production, suggesting that the work function for aluminum with adsorbed water and other adsorbates could be as low as 3.3 - 3.5 eV (375 - 355 nm). However, when using UV light above $\sim$ 345 nm to stimulate photoelectrons, the electron production degraded quickly with time as the UV light desorbed the water and other absorbents from the aluminum surface. 

We also studied the dependence of photoelectron production due to the effect of water and adsorbates on the surface of a magnesium alloy (AZ81 (8\% Al, 1\% Zn)) cathode.  The output photoelectron production is significantly less for the magnesium cathode when compared to the Al cathode, suggesting that the water adsorbate does not shift the work function to a value as low as for the aluminum.  The work function of oxidized Mg can be up to 1 eV less than the work function for the bare metal surface which is 3.5 - 3.7 eV (345 - 330 nm).\cite{MgOxide2}  We were, however, unable to stimulate photoelectrons with a 370 nm UV LED from a Mg cathode, whereas Al surfaces with adsorbates consistently produced a photoelectron signal.  Our studies suggest that efficiency of photoelectron stimulation from Mg surfaces depends on the preparation process of the cathode material and is more sensitive than Al surfaces.

\subsection{Magnetic Field Cancellation}

Two sets of Helmholtz coils are mounted external to the vacuum chamber and are used to cancel the influence of the Earth's magnetic field on electron trajectories. The electron beam direction lies along the east-west vector, which is the weakest component of the Earth's magnetic field, $B_{\rm{x}}$. The north-south magnetic field component ($B_{\rm{y}} \sim$ 22,000 nT) is compensated using 1.5 m diameter coils whose centers lie $\sim$0.5 m from the center of the chamber. The vertical component ($B_{\rm{x}} \sim$ 44,000 nT) of the field is compensated using 1.2 m diameter coils located 0.5 m above and below the chamber center, respectively. Each coil has five turns, and together each pair are run in Helmholtz configuration with 5 A of current. In total, these coils produce a field which is calculated to spatially vary by less than 10\% (or 3, 500 nT) over a 0.125 m$^3$ region. 

% #########################################################################
\section{Measurements and discussion}

The maximum possible value of the energy spread $\Delta E_{\rm{LON}}$ of the electron beam in the longitudinal direction (along the beam axis) assumes that the work function is zero and therefore equals the incident photon energy, e.g., $\Delta E_{\rm{LON}}$ $<$ 3.4 eV for a 365 nm photon. The maximum possible value for the latitudinal energy spread $\Delta E_{\rm{LAT}}$ corresponds to twice the maximum value of $\Delta E_{\rm{LON}}$ because photoelectrons can be emitted from the cathode in opposite longitudinal directions, both perpendicular to the beam axis. For a 365 nm photon, the maximum value of $\Delta E_{\rm{LAT}}$ is therefore 6.8 eV. We expect that the actual thermal energy of the beam exiting the photocathode is smaller because a work function of zero is unrealistic and the maximum kinetic energy of an electron emerging from the cathode is the difference between the work function and the incident UV photon energy. Conversely, the fringe fields at the grid in the grid plate (Fig.\;\ref{fig:Schematic}) can slightly deflect some electrons in the latitudinal direction, converting some longitudinal energy into latitudinal energy, increasing the energy spread in both the longitudinal and latitudinal directions.

The energy width of the output electron beam has been measured using a single grid located in front of the CEM detector that acts as a simplified retarding potential analyzer (RPA). The count rate of electrons detected by the CEM is monitored as voltage applied to the RPA grid is varied across a range $V_0 \pm 70$ V where $V_0$ is the voltage applied to the cathode and, therefore, equal to the magnitude of the accelerating potential of the electron beam.

\begin{figure}[htp]
\begin{center}
\includegraphics[width=4in,height=2.57in]{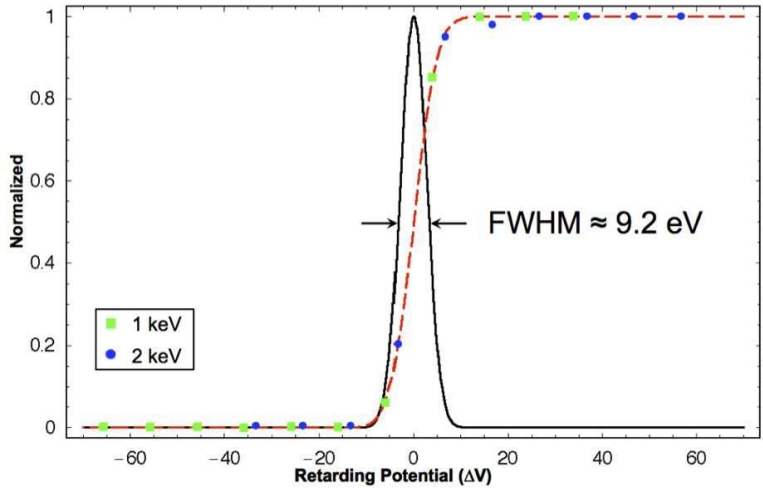}
\caption{(color online) The combined energy width of 1 keV and 2 keV electron beams using 365 nm UV LED was measured by varying the voltage across a single retarding potential analyzer grid in front of the CEM detector.  The red (dashed) line represents an error function ($\frac{2}{\sqrt\pi} \int_{V_1}^{V_2} e^{-x^2} dx$) fit to the data.  The black (solid) line is a derivative of the error function and reveals the energy width of the beam derived using the fitted error fnction.}
\label{fig:EnergyWidth}
\end{center}
\end{figure}

Electrons with an energy less than the RPA grid voltage cannot be detected, whereas electrons with energy greater than the RPA grid voltage can pass through the grid and be detected. The data points in Fig.\;\ref{fig:EnergyWidth} are the normalized count rate of the CEM detector as a function of the voltage applied to the RPA grid for incident electron beams at energies 1 and 2 keV. An error function was fit to the data and is plotted as the red line in the figure. The corresponding Gaussian distribution (the derivative of the error function) is plotted as the black line with the derived full width at half maximum (FWHM) value of 9.2 eV. This provides an upper limit to the energy spread of the beam, primarily in the longitudinal direction. Although the RPA grid generally acts in the longitudinal direction, the latitudinal energy spread can broaden the energy width due to non-longitudinal fringing fields at the RPA grid that can strongly influence electron trajectories as they are slowed to very low energies at the RPA grid. 

\begin{figure}[htp]
\begin{center}
\includegraphics[width=5.5in,height=2.88in]{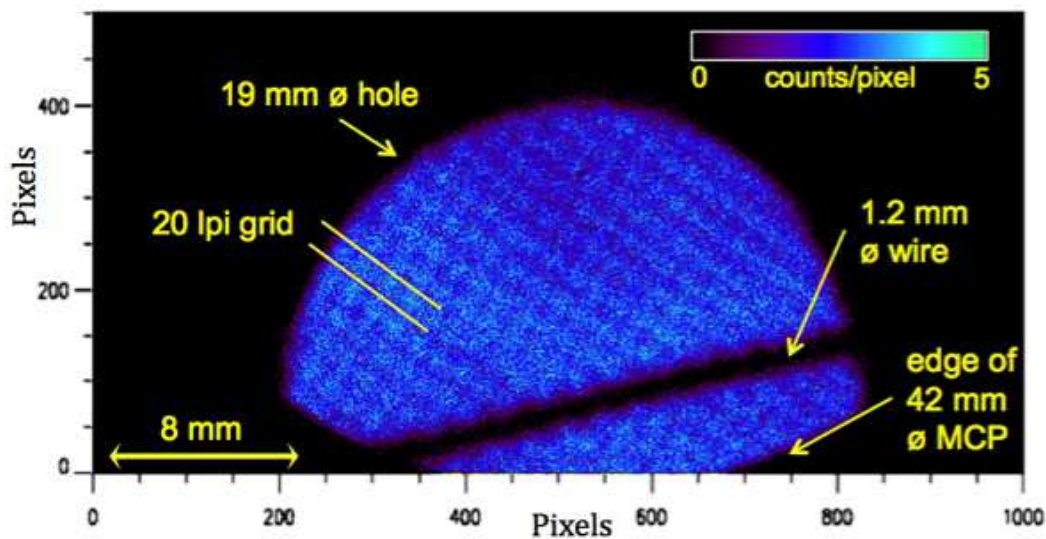}
\caption{(color online) An image derived using the position-sensitive MCP detector of a 5 keV electron beam with an aperture is shown on the left. The MCP was placed 100 mm away from the grounded grid of the electron source, which used a 365 nm LED. The grounded grid (20 lpi) of the electrons source grid plate, a 1.2 mm diameter wire and a 19 mm hole are imaged onto the MCP. Analysis of this image shows that the divergence of a 5 keV electron beam is less than $1^{\circ}$.}
\label{fig:MCPImage}
\end{center}
\end{figure}

The spatial profile of the beam and the angular divergence of a 5 keV electron beam were measured using a 42 mm diameter imaging microchannel plate (MCP) detector with $65\;\mu$m resolution.  Fig.\;\ref{fig:MCPImage} shows an image of an nominal 50 mm diameter electron beam. The 20 lpi ground grid of the electron source, located 100 mm from the MCP detector, is clearly observed as the orthogonal grid lines of lower counts in the image. A 1.2 mm diameter wire and 19 mm aperture were additionally placed 100 mm from the MCP detector. For this image we used a single 365 nm UV LED operating at 10 mA without a lens. Analysis of the images of the wire and aperture show an angular HWHM divergence of  $<1^{\circ}$ , which corresponds to a FWHM latitudinal energy spread of $\sim$ 3 eV, consistent with our previous expectations and results.

%On the right is an MCP image of the electron beam taken without apertures.

\begin{figure}[htp]
\begin{center}
\includegraphics[width=6.5in,height=2.21in]{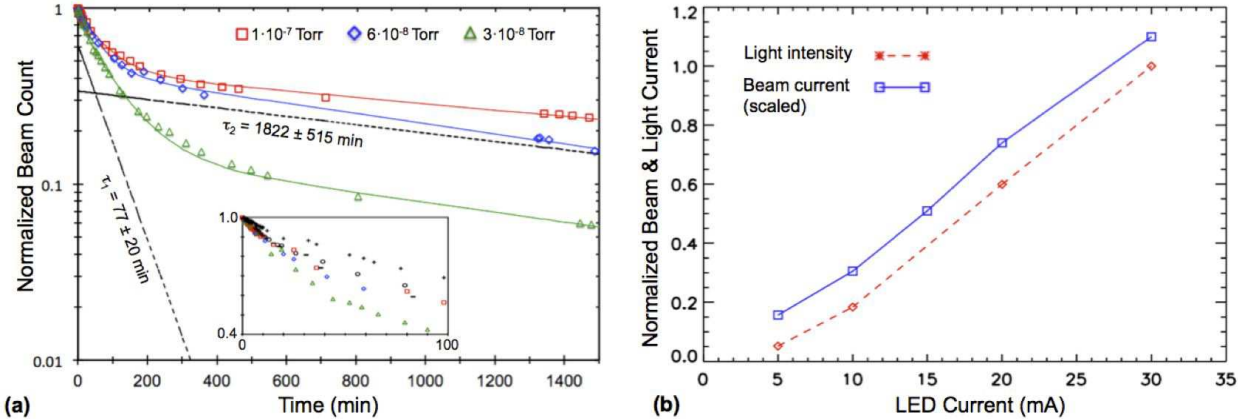}
\caption{(color online) (a) Normalized count rate (or beam current) as a function of time for three runs at pressures, $1\cdot10^{-7}$, $6\cdot10^{-8}$, and $3\cdot10^{-8}$ Torr. The cathode is aluminum and the LED is 260 nm.  The average of the data is well-represented by a double exponential fit  $0.62(7) \cdot e^{-\rm{t}/77(20)}+0.34(8) \cdot e^{-\rm{t}/1822(515)}$, which are shown separately as dashed lines. We interpret this behavior as a system predominantly driven by two adsorbate species whose characteristic desorption curves are each shown as the dashed black lines. The inset shows that the cathode surface conditions fully recover after venting to atmosphere or turning off the LED for 12 hours.  Additional short runs are included to show reproducible recovery. (b) Normalized light intensity and beam current as a function of LED current.   The electron beam current and LED light intensity are normalized to unity at an input LED current of 30 mA.  Beam current is offset for clarity.}
\label{fig:Stability}
\end{center}
\end{figure}

We additionally measured the stability of the electron beam current of a 10 keV electron beam over time. Fig.\;\ref{fig:Stability}(a) is representative of results at multiple energies for a polished aluminum cathode stimulated by a 260 nm LED at three different vacuum pressures: $1 \cdot 10^{-7}$, $6 \cdot 10^{-8}$, and $3\cdot 10^{-8}$ Torr. A current-regulated power supply provided a constant 20 mA of current to the LED, and a photodetector monitored the light output of the LED, as shown in Fig.\;\ref{fig:Photo}. The beam current was measured by a channel electron multiplier. Three runs were made lasting 1450 minutes and three runs were made lasting less than 200 minutes. Each run was separated by either a break in vacuum in which the system was vented to atmosphere or a minimum of 12 hours in which the photodiode was turned off. Under both of these conditions, ambient volatile species are allowed sufficient time to reach quiescent equilibrium surface concentrations and recovery to the initial work function until perturbed again by exposure to UV light from the LEDs. The average of the data show that, immediately after the UV LED is turned on, the electron source beam current steadily drops by about 1.3\%/minute for the first 20 minutes. After $\sim$ 20 minutes, the decrease in beam current begins to slow and after 400 minutes the beam current drops to 0.05\%/minute, as shown in Fig.\;\ref{fig:Stability}(a). The inset of Fig.\;\ref{fig:Stability} show that when the LED is turned off for several hours the beam current recovers to within 5\% of the original value, showing nearly complete recovery to the initial surface conditions.

The reduction of beam current with increasing duration of UV exposure from the LED is a result of the desorption of adsorbates from the cathode surface. In particular, adsorbed water has been shown to decrease the work function of surfaces, and the decrease in electron beam current with increasing UV exposure is characteristic of an increasing work function.  RGA analysis shows that water is a dominant residual gas in the vacuum, so a reduction in beam current with increasing time of exposure is both expected and observed.  The desorption of water initially dominates the rate of change of the electron beam current.   As the rate of water desorption decreases, the desorption of other adsorbates, mainly hydrocarbons like vacuum pump oils, begins to determine the rate of change of the electron beam current.

If we assume the simple case of UV-induced desorption without replenishment, the rate of desorption from an incident areal UV flux $\Phi_{\rm{UV}}$ is
\hspace{-10 mm}
\begin{equation}
\frac{dn_{\rm{A}}}{dt} = - n_{\rm{A}}\sigma_{\rm{D}}\Phi_{\rm{UV}}
\label{eqn:1}
\end{equation} 
where $n_{\rm{A}}$ is the areal density of the adsorbate on the surface and $\sigma_{\rm{D}}$ is cross section of desorption by a UV photon. The integrated equation is
\hspace{-10 mm}
\begin{equation}
n_{\rm{A}}(t) = n_{\rm{A}}(0)\;\rm{exp}(-t/\tau) 
\label{eqn:2}
\end{equation} 
such that the adsorbate coverage depletes exponentially with a characteristic time constant of $\tau = (\sigma_{\rm{D}} \Phi_{\rm{UV}})^{-1}$. We find that the decrease in the output electron beam current is well-represented by the double exponential function 
\hspace{-10 mm}
\begin{equation}
n_{\rm{A}}(t) = 0.62(7)\;\rm{exp}(-t/77(20)) + 0.34(8)\;\rm{exp}(-t/1822(515)) 
\label{eqn:3}
\end{equation} 
where $t$ is in minutes and standard deviations are provided in parenthesis. The two exponentials are shown as dashed lines in Fig.\;\ref{fig:Stability}(a). The fit suggests that there are two dominant adsorbate species that influence the work function of the cathode surface. When present, both of these adsorbates act to decrease the work function.

Furthermore, we do not find that the beam current approaches a constant, equilibrium value over the time period of the experiment that is indicative of adsorbate replenishment. However, we do observe complete replenishment when the UV LED is turned off for 12 or more hours. This suggests that the replenishment of adsorbates is inhibited by the UV illumination and consistent with our Eqn.\;\ref{eqn:1} assumption of desorption without replenishment.

Fig.\;\ref{fig:Stability}(b) shows the electron beam current and the LED output light intensity measured as a function of LED current for a 5 keV electron beam for cathode conditions that have fully recovered. The beam current at maximum LED current (30 mA) for this particular measurement was  $\sim 10^5$ electrons cm$^{-2}$ s$^{-1}$. For short time periods ($< 5$ minutes), the surface condition changes minimally (Fig.\;\ref{fig:Stability}(a)) and these results are reproducible to within $1.5\%$. Both the decrease in electron beam current with increasing exposure to UV and the correlation between electron beam current and input current to the UV LED are predictable and reproducible. By continuously or intermittently sampling the electron beam output, the beam current can be monitored. Feedback for active control of the LED current could additionally be incorporated to maintain a constant beam current for long periods of time.

Since UV LED's have comparatively short rise times of $1\;\mu$s or less, this type of electron source can have robust amplitude modulation capabilities. Using arbitrary waveform generators to control the current of the LED, electron pulses, well-defined in amplitude and time, can be tailored to specific needs, for example to assess a spectrometer's response to fast, dynamic processes in space plasmas.

\newpage
%\vspace{100 mm}
%{\color{white}hi\\

% #########################################################################
\section{Summary}

We have built and demonstrated a simple, inexpensive, and compact 0.1-30 keV electron source using a UV LED to generate photoelectrons from a large-area photocathode and acceleration perpendicular to the photocathode using a simple, proximity-focusing grid. We have demonstrated performance using polished Mg and Al photocathodes at pressures of $\sim 10^{-7}$ torr. Both Al and Mg form a native oxide surface, and this surface is substantially covered with adsorbates at this vacuum pressure. The high output electron current obtained from these LEDs is enabled by their generation of UV photons of sufficient energy to generate photoelectrons that overcome the work function associated with the native oxide surfaces covered with adsorbates. Its simplicity allows it to be implemented for a variety of applications and includes applications that require fast, controllable amplitude modulation of the electron beam current. The low resources and versatile design incorporated into the construction and operation of this electron source may also  enable precision, in situ control of spacecraft potential, as well as a stimulus for a low intensity photon bremsstrahlung x-ray sources.

\begin{acknowledgments}
Work supported by Los Alamos National Laboratory (Department of Energy; LA-UR 11-04716) and NASA (RBSP Mission).
\end{acknowledgments}

% #########################################################################

%\newpage
%\vspace{100 mm}
%{\color{white}hi\\

\vspace{200 mm}


\begin{thebibliography}{10}

\bibitem{SSD1}
H.O. Funsten, D.M. Suszcynsky, S.M. Ritzau, and R.~Korde.
%\newblock {Response of 100\% internal quantum efficiency silicon photodiodes to 200 eV - 40 keV electrons}.
\newblock {\em IEEE Trans. Nucl. Sci.}, 44:2561, 1997.

\bibitem{SSD2}
C.S. Tindall, N.P. Palaio, B.A. Ludewigt, S.E. Holland, D.E. Larson, D.W. Curtis, S.E. McBride, T.~Moreau, R.P. Lin, and V.~Angelopoulos.
%\newblock {Silicon detectors for low energy particle detection}.
\newblock {\em IEEE Trans. Nucl. Sci.}, 55:797, 2008.

\bibitem{Plasma0}
M.~Lessard, G.P. Garbe, R.L. Arnoldy
%\newblock {A calibration source for low energy electron detectors}.
\newblock {\em Measurement Techniques in Space Plasmas: Particles,}
\newblock {R.~Pfaff, J.~Borovsky, and D.T. Young, Eds.,}
\newblock {Geophys. Monogr. Ser.}, 102, 301, 1998.

\bibitem{Plasma1}
D.T. Young, et al.
%\newblock {Cassini Plasma Spectrometer Investigation}.
\newblock {\em Space Sci Rev}, 114:1, 2009.

\bibitem{Plasma2}
C.W. Carlson, J.P. McFadden, P.~Turin, D.W. Curtis, and A.~Magoncelli.
%\newblock {The electron and ion plasma experiment for FAST}.
\newblock {\em Space Sci Rev}, 98:33, 2001.

\bibitem{Katrin1}
K.~Valerius, M.~Beck, H.~Arlinghaus1, J.~Bonn, V.M. Hannen, H.~Hein, B.~Ostrick, S.~Streubel, Ch.~Weinheimer, and M.~Zbo$\check{\text{r}}$il.
%\newblock {A UV LED-based fast-pulsed photoelectron source for time-of-flight studies}.
\newblock {\em New J. Physics}, 11:063018, 2009.

\bibitem{LISA1}
K.~Sun, B.~Allard, S.~Buchman, S.~Williams and R.~L.~Byer.
%\newblock {LED deep UV source for charge management of gravitational reference sensors}.
\newblock {\em Class. Quantum Grav.}, 23:S141, 2006.

\bibitem{LISA2}
S.E. Pollack, M.D. Turner, S.~Schlamminger, C.A. Hagedorn, and J.H. Gundlach.
%\newblock {Charge management for gravitational-wave observatories using UV LEDs}.
\newblock {\em Phys. Rev. D}, 81:021101(R), 2010.

\bibitem{Kimball}
To our knowledge, no commercially available electron sources can provide a 0.1 - 30 keV energy range and low beam current ($<10$\;nA).  Kimball Physics produces a number of models that are close, like EMG-4103 / EGPS-4103 Electron Gun System: 1keV - 30 keV, 10 nA to 1 $\mu$A, spot size of 15 to 500 $\mu$m.
\newblock {Kimball Physics, Wilton, NH}.

\bibitem{CNT}
Y.~Hozumi.
%\newblock {Development of electron gun of carbon nanotube cathode}.
\newblock {\em Proceedings of the 2005 Particle Accelerator Conference, Knoxville, Tennessee. IEEE 2005}, p. 1392-1394, 2005.

\bibitem{Katrin2}
K.~Hugenberg.
%\newblock {An angular resolved pulsed UV LED photoelectron source for KATRIN}.
\newblock {\em Prog. Part. Nucl. Phys.} 64:288, 2010.

\bibitem{Yamamoto}
S.~Yamamoto.
%\newblock {Fundamental physics of vacuum electron sources}.
\newblock {\em Rep. Prog. Phys.} 69:181, 2006.

\bibitem{WFAl00}  % WF Experimental value - Al
J.K. Grepstad, P.O Gartland, and BJ. Slagsvold.
%\newblock {Anisotropic work function of clean and smooth low index faces of aluminum}.
\newblock {\em Surf. Sci.}, 57:348, 1976.

\bibitem{AlOxide0}  %Stability of oxide layer and log growth  0.1 - 0.4 eV  (depending on pressure and time) UV exposure complicates
V.K. Agarwala and T.~Fort, Jr.
%\newblock {Nature of the stable oxide layer formed on an aluminum surface by work function measurements}.
\newblock {\em Surf. Sci.}, 54:60, 1976.

\bibitem{UVDesorp}  %Ice desorption from UV, but contains lots of details over broad energy range about desorption mechanism
M. S. Westley, R.A. Baragiola, R.E. Johnson and G.A. Baratta
%\newblock {Ultraviolet photodesorption from water ice}.
\newblock {\em Planet Space Sci.}, 43:1311, 1995.

\bibitem{AlOxideWater0}  %Pressure vs. time equilibrium for water adsorption  -First report.
J.A. Ramsey.
%\newblock {The emission of electrons from aluminum abraded in atmospheres of air, oxygen, nitrogen and water vapour}.
\newblock {\em Surf. Sci.}, 8:313, 1967.

\bibitem{AlOxideWater1}  %Pressure vs. time equilibrium for water adsorption  -- I think this is the one.  1.4eV change
T.~Fort Jr. and R.L. Wells.
%\newblock {Adsorption of water on clean aluminum by measurement of work function changes}.
\newblock {\em Surf. Sci.}, 32:543, 1972.

\bibitem{MgOxide2}  % predicts about 1.2 eV, which is far less than we see ~ 450 nm
P.~Lange, D.~Grider, H.~Neff, J.K. Sass and R.~Unwin.
%\newblock {Limitations of the Fowler method in photoelectric work function determination: oxygen on magnesium single crystal surfaces}.
\newblock {\em Surf. Sci.}, 118:L257, 1982.






% EXTRA Citations


%\bibitem{Source1}
%G.~Paschmann, E.G. Shelley, C.R. Chappell, R.D. Sharp, and L.F Smith.
%\newblock {Absolute efficiency measurements for channel electron multipliers utilizing a unique electron source}.
%\newblock {\em Rev. Sci. Instrum.}, 23:S141, 1970.


%\bibitem{Pulsed1}
%D.~Wytrykus, M.~Centurion, P.~Reckenthaeler, F.~Krausz, A.~Apolonski, E.~Fill.
%\newblock {Ultrashort pulse electron gun with a MHz repetition rate}.
%\newblock {\em Appl. Physics B}, 96:309, 2009.

%\bibitem{SpekCalc}
%G.G. Poludniowskia and P.M. Evans.
%\newblock {Calculation of x-ray spectra emerging from an x-ray tube. Part I. Electron penetration characteristics in x-ray targets}.
%\newblock {\em Med. Phys.}, 34(6):2164, 2007.



%\bibitem{AlOxide1}
%E.E. Huber, Jr. and C.T. Kirk, Jr.
%\newblock {Work function changes due to the chemisorption of water and oxygen on aluminum}.
%\newblock {\em Surf. Sci.}, 5:447, 1966.

%\bibitem{AlOxide2}  %Pressure vs. time equilibrium
%V.K. Agarwala and T.~Fort, Jr.
%\newblock {Work function changes during low pressure oxidation of aluminum at room temperature}.
%\newblock {\em Surf. Sci.}, 45:470, 1974.


%\bibitem{AlOxideWater1}  %Pressure vs. time equilibrium for water adsorption  -- I think this is the one.  1.4eV change
%T.~Fort Jr. and R.L. Wells.
%\newblock {Adsorption of water on clean aluminum by measurement of work function changes}.
%\newblock {\em Surf. Sci.}, 32:543, 1972.

%\bibitem{AlOxideWater2}  %Pressure vs. time equilibrium for water adsorption  -- A review.
%V.K. Agarwala and T.~Fort, Jr.
%\newblock {Effect of pressure and temperature on changes in the work function of aluminum during interaction with oxygen.}.
%\newblock {\em Surf. Sci.}, 48:527, 1975.

%\bibitem{AuWater1}  %Pressure vs. time equilibrium for water adsorption  -- possibly relevant for explanation of Ti, Mg
%R.L. Wells and T.~Fort.
%\newblock {Adsorption of water on clean gold by measurement of work function changes}.
%\newblock {\em Surf. Sci.}, 32:554, 1972.


%\bibitem{WFAl0}
%C.J. Fall, N.~Binggeli, and A.~Baldereschi.
%\newblock {\em Anomaly in the anisotropy of the aluminum work function},
%\newblock {\em Phys. Rev. B}, 58:7544, 1998.

%\bibitem{WFAl1}  % Not really relevant
%M.W. Roberts and B.R. Wells.
%\newblock {Chemsiorption of oxygen by aluminum}.
%\newblock {\em Surf. Sci.}, 15:325, 1969.

%\bibitem{WFAl2}
%R.L. Wells and T.~Fort, Jr.
%\newblock {Interaction of oxygen with clean aluminum surfaces by measurement of work function changes}.
%\newblock {\em Surf. Sci.}, 33:172, 1972.

%\bibitem{WF0}
%D.R. Lide, Ed.
%\newblock {\em CRC Handbook of Chemistry and Physics},
%\newblock {(CRC Press/Taylor and Francis, Boca Raton Fl 2009), 89th ed}.

%\bibitem{WF1}
%C.S. Beleznai, D.~Vouagner, J.P. Girardeau-Montaut.
%\newblock {Work function variation during UV laser-induced oxide removal}.
%\newblock {\em Appl. Surf. Sci.}, 138:6, 1999.

%\bibitem{MgOxide1}
%T.F. Gesell and E.T. Arakawa.
%\newblock {Work function changes during oxygen chemisorption on fresh magnesium surfaces}.
%\newblock {\em Surf. Sci.}, 33:419, 1972.


%\bibitem{Oxide4}
%R.~Jaeckel and B.~Wagner.
%\newblock {Photoelectrlc measurement of the work function of metals and its alteration after gas adsorption}.
%\newblock {\em Vacuum}, 13:509, 1963.

%\bibitem{Chemsorp1}  %Chemisorption of water...equations for equilibrium are contained within...can shed light onto equilibrium of situation.
%S.~Khoobiar, J.L. Carter, and P.J. Lucchesi.
%\newblock {The electronic properties of aluminum oxide and the chemisorption of water, hydrogen, and oxygen}.
%\newblock {\em J. Phys. Chem.}, 75:1682, 1968.


%\bibitem [{Note1()}]{Note1}%
%  \BibitemOpen
%  \bibinfo {note} {Note information.}\BibitemShut {Stop}%


%\bibitem [{Note2()}]{Note2}%
%  \BibitemOpen
%  \bibinfo {note} {{{\color{blue}Different grids , e.g., 70 lpi nickel, and non-conductive Kapton tape were used in other design models and no effects were noticed.}}}\BibitemShut {Stop}%


\end{thebibliography}
\end{document}